\RequirePackage{fix-cm}
\documentclass{svjour3}                     
\smartqed  
\usepackage{graphicx}
\journalname{Gen. Relativ. Grav.}

\begin{document}

\title{A class of conformally flat solutions for systems undergoing radiative gravitational collapse}

\titlerunning{A class of conformally flat solutions}        

\author{Ranjan Sharma \and Shyam Das \and Ramesh Tikekar}

\authorrunning{Sharma {\em et al}} 

\institute{Ranjan Sharma \and Shyam Das \at P. D. Women's College, Jalpaiguri 735101, India.\\
          \email{rsharma@iucaa.ernet.in}       
          \and
           Ramesh Tikekar \at Department of Mathematics, Sardar Patel University, Vallabh Vidyanagar, Gujarat, India.}

\date{Received: date / Accepted: date}

\maketitle

\begin{abstract}
We present a class of conformally flat solutions of the Einstein's field equations for spherical systems undergoing gravitational collapse accompanied with radial heat flux. The interior space-time of the collapsing matter is chosen to be conformal to the Robertson-Walker space-time so that its Weyl tensor vanishes identically. The conditions which ensure the smooth matching of the interior space-time of the collapsing system, across its boundary, with the exterior space-time of a radiating star which is described by Vaidya metric lead to an equation governing its overall subsequent evolution. New solutions of this equation have been shown to provide physically viable models of collapsing stars.
\end{abstract}

\section{\label{sec1}Introduction}  
Study of relativistic models of gravitationally collapsing systems has been receiving considerable attention since Oppenheimer and Snyder\cite{OppenS} first examined the collapse of a homogeneous spherical distribution of matter in the form of dust. It is known that a star which loses its equilibrium, collapses under self gravity, ultimately culminating in the state of space-time singularity. The state of singularity may be a naked one, in the sense that light rays emanating from it can reach any far-away observer or a black hole implying that the light rays from it may be trapped within a surface called its event horizon. Our understanding about the final outcome of any collapsing stellar configuration is known to rely on a couple of adopted conjectures only in this regard \cite{Joshibook,Thornebook}.  Accordingly, studies of realistic models of gravitationally collapsing systems are expected to play a key role in promoting our understanding of the nature and dynamics of the collapsing stellar systems.  Many studies aimed at examining various aspects of gravitationally collapsing stellar systems of different kinds of matter distributions have been carried out on varying types of space-time background.
The space-time in the exterior region of a radiating fluid sphere is known to be described by Vaidya\cite{Vaidya} metric. The formulation of the boundary conditions smoothly joining the interior space-time-metric of the collapsing matter across its boundary with the appropriate form of Vaidya metric of the exterior space-time, as first proposed by Santos\cite{Santos} has given a tremendous impetuous for studies in this direction.

In the present work, we have proposed a relativistic model of a spherically symmetric matter source, whose collapse is accompanied with dissipation in the form of radial heat flux following Santos\cite{Santos}. The background space-time metric is chosen so as to have vanishing Weyl tensor implying its conformal flatness. Conformally flat space-times, in the context of radiating fluid spheres, were first studied by Som and Santos\cite{Som}. Later, the most general class of conformally flat solutions for a shear-free radiating star were obtained and examined by Maiti\cite{Maiti}, Modak\cite{Modak}, Banerjee {\em et al}\cite{Baner}, Patel and Tikekar\cite{Patel}, Sch$\ddot{a}$fer and Goenner\cite{Dirk} and Ivanov\cite{Ivanov}. Herrera {\em et al}\cite{Herrera1} have critically examined models of shear-free collapsing fluids accompanied with dissipation of heat on the space-time background subject to the constraint that the associated Weyl tensor should vanish. In the present approach, the geometry of the background space-time of the non-adiabatically collapsing matter is chosen to be conformal to that of the space-time of Robertson-Walker metric so that the Weyl tensor is known to vanish since the Robertson-Walker space-time is already conformally flat.

The paper has been organized as follows. In Sec.~\ref{sec2}, the Einstein  field equations for non-adiabatically collapsing spherical matter distributions with radial heat flow have been formulated on the conformally flat space-time background with pre-assigned metric form. The pressure isotropy relation in this approach is found to lead to two classes of solutions for the system. The junction conditions which smoothly join the interior space-time with the  exterior Vaidya\cite{Vaidya} metric across the boundary surface $\Sigma$ have been obtained in Sec.~\ref{sec3}. In Sec.~\ref{sec4}, the explicit expressions for the physical quantities and the constraints on the model-parameters in view of physical requirements have been obtained. In Sec.~\ref{sec5}, specific models have been examined by considering a couple of solutions of the equation governing the evolution of the collapse and their physical viability is shown using graphical methods. Sec.~\ref{sec6} contains a discussion accompanied with some concluding remarks.

\section{\label{sec2} Conformally flat radiating collapse model}
We begin with the space-time describing the interior of a spherically symmetric collapsing matter fluid with radial heat flux with its metric
\begin{equation}
ds_{-}^2 = \frac{1}{A^2(r,t)}\left[dt^2 - \frac{dr^2}{1-kr^2} - r^2(d\theta^2 + \sin^2\theta d\phi^2)\right].\label{intm1}
\end{equation}
The evolution with time of the configuration is governed by the function  $A(r,t)$ determined by relativistic field equations. The Weyl tensor for the space-time of metric (\ref{intm1}) will vanish indicating its conformal flatness since the space-time is also conformal to the Robertson-Walker metric which follows on setting $A(r,t)= a(t)$. 

The matter content of the collapsing object in the presence of heat flux is described by the energy momentum tensor
\begin{equation}
T_i^j = (\rho + p)u_i u^j - p \delta_i^j + q_i u^j + q^j u_i,\label{emt1}
\end{equation}
where $\rho$ is the energy density, $p$ is the isotropic fluid pressure, $ u^{i} = (A, 0, 0, 0)$ is the $4$-velocity of the
fluid and $q^{i} = (0, q, 0, 0)$ is the radially directed heat flux vector. 

Subsequently, Einstein field equations relating metric potentials of (\ref{intm1}) with the dynamical variables of (\ref{emt1}) is a system consisting of the following four equations:
\begin{eqnarray}
\rho &=& 3k A^2+3{\dot{A}}^2-3(1-k r^2)A'^2+2(1-k r^2)A A'' +\frac{2}{r}(2-3 k r^2)A A', \label{Eq1}\\
p &=& - k A^2 + 2 A\ddot{A}-3{\dot{A}}^2+3(1-k r^2){A'}^2-\frac{4}{r}(1-k r^2)A A',  \label{Eq2}\\
p &=& -kA^2+2A\ddot{A}-3{\dot{A}}^2+3(1-kr^2){A'}^2-2(1-kr^2)AA''\nonumber\\
&&-\frac{2}{r}(1-2k r^2)AA',\label{Eq3}\\
q &=& -2(1-kr^2)A^2\dot{A}',\label{Eq4}
\end{eqnarray}
where an overhead prime ($'$) and an overhead dot ($.$) denote differentiations with respect to $r$ and $t$, respectively.

Eqs.~(\ref{Eq2}) and (\ref{Eq3}) lead to
\begin{equation}
\frac{A''}{A} -\frac{A'}{rA}\frac{1}{(1-kr^2)}= 0.\label{pi}
\end{equation}
Eq.~(\ref{pi}) usually referred to as pressure isotropy condition is easily integrable and admits two solutions:

{\bf Case I:} $k=0$:
\begin{equation}
A(r,t) = \xi(t)r^2+\eta(t).\label{sol1}
\end{equation} 

{\bf Case II:} $k \neq 0$:
\begin{equation}
A(r,t) = \xi(t)\sqrt{1-k r^2} +\eta(t).\label{sol2}
\end{equation} 
On setting $\xi =0$ one obtains the FRW model of the universe while on setting $\xi=$ a constant, one obtains a perfect fluid space-time solution without any dissipation or heat flow.  Accordingly, solutions obtained on setting time dependence of the functions $\xi(t)$ and $\eta(t)$ appropriately will lead to models of collapsing fluid systems in the presence of heat flux.

Note that, in the context of shear-free radiating collapse, the conformally flat space-time metric obtained earlier by Herrera {\em et al}\cite{Herrera1} has the form (in coordinates ($\tilde{t}, \tilde{r}, \theta, \phi$))
\begin{equation}
ds_{-}^2 = \frac{[\tilde{\chi}(\tilde{t})\tilde{r}^2+1]^2}{[\tilde{\xi}(\tilde{t})\tilde{r}^2+\tilde{\eta}(\tilde{t})]^2} d\tilde{t}^2 - \frac{1}{[\tilde{\xi}(\tilde{t})\tilde{r}^2+\tilde{\eta}(\tilde{t})]^2}\left[d\tilde{r}^2 + \tilde{r}^2(d\theta^2 + \sin^2\theta d\phi^2\right].\label{Herrm1}
\end{equation} 
Obviously, the solution (\ref{sol1}) is a sub-class of the Herrera {\em et al}\cite{Herrera1} model which can be obtained by setting $\tilde{\chi}(\tilde{t}) = 0$ in Eq.~(\ref{Herrm1}). Further, if we set $\tilde{\chi} = k$ and make the following coordinate transformations
$$\tilde{r} = \frac{r}{(1+\sqrt{1-k r^2})},~~~~~~\tilde{t} = \frac{t}{2},$$
the solution (\ref{sol2}) can be obtained from (\ref{Herrm1}) by identifying the parameters $\eta = \tilde{\eta}+\tilde{\xi}/k$ and $\xi = \tilde{\eta}-\tilde{\xi}/k$. Hence, the solution (\ref{sol2}) also turns out to be a sub-case $\tilde{\chi}=$ a constant of Eq.~(\ref{Herrm1}). Since all conformally flat space-times are conformally related, this is expected.

It should be stressed here that even though the classes of solutions obtained in this paper are special cases of the conformally flat solution of \cite{Herrera1}, the solutions have been obtained by adopting a different approach. The conformally flat solution of \cite{Herrera1} has been obtained by equating the Weyl tensor of the associated space-time to zero while the solutions reported in this paper have been obtained by imposing the pressure isotropy condition in a space-time with geometry conformal to a conformally flat Robertson-Walker space-time. The space-time geometry of the non-adiabatically collapsing fluid sphere continues to be conformal to the homogeneous geometry of the RW space-time as the collapse proceeds. In addition, the two classes of solutions have the following distinctive features. If $\tilde{\xi}(\tilde{t}) = (k/4)\tilde{\eta}(\tilde{t})$, $\tilde{\chi}(\tilde{t}) = (k/4)$, on choosing a new time coordinate $dT  = d\tilde{t}/\tilde{\eta}(\tilde{t})$, one finds that the space-time of (\ref{Herrm1}) corresponds to that of a FRW fluid sphere without heat flux. In our model, the space-time degenerates to that of a FRW fluid sphere without heat flux when $\xi(t) = 0$. It is noteworthy that our choice of curvature coordinates brings out to attention the role of the curvature parameter $k$ of RW space-time explicitly in the evolution of the collapse.

\section{\label{sec3}Junction conditions}

The space-time exterior to the collapsing matter source will be filled with radiation and is appropriately described by the Vaidya\cite{Vaidya} metric for a radiating star
\begin{equation}
ds_{+}^2 = \left(1-\frac{2m(v)}{{\sf r}}\right)dv^2 + 2dv d{\sf r} - {\sf r}^2[d \theta^2 + \sin^2 \theta d\phi^2]. \label{Vm}
\end{equation}

The interior space-time (\ref{intm1}) should smoothly be joined with that of the Vaidya metric (\ref{Vm}) across the boundary surface $\Sigma$ separating the interior ($r \leq r_\Sigma$) and the exterior ($ r \geq r_\Sigma$) space times. Following Santos\cite{Santos}, we shall stipulate the boundary conditions which ensure continuity of (i) the metrics of the interior and exterior space times with that on $\Sigma$, and (ii) the extrinsic and intrinsic curvatures of $\Sigma$.

Let $g_{ij}$ be the intrinsic metric to $\Sigma$ such that 
\begin{equation}
ds^2_{\Sigma} = g_{ij}d\xi^i d\xi^j,~~~~~i=1,2,3\label{intrinm1}
\end{equation} 
and $g^\pm_{\alpha\beta}$ be the metric corresponding to the exterior ($+$)/interior ($-$) regions so that
\begin{equation}
ds^2_\pm = g^\pm_{\alpha\beta}d\chi^\alpha_\pm d\chi^\beta_\pm,~~~~~\alpha=0,1,2,3.\label{intrinm2}
\end{equation}
Then the first condition implies
\begin{equation}
(ds^2_{-})_\Sigma = (ds^2_{+})_\Sigma = ds^2_\Sigma.\label{eqj1} 
\end{equation}
The second condition implies
\begin{equation}
(K^+_{ij})_{\Sigma} = (K^-_{ij})_{\Sigma},\label{eqj2}
\end{equation}
where,
\begin{equation}
K^\pm_{ij} \equiv -n^\pm_\alpha\frac{\partial^2\chi^\alpha_\pm}{\partial \xi^i\partial \xi^j}-n^\pm_\alpha\Gamma^\alpha_{\mu\nu}\frac{\partial\chi^\mu_\pm}{\partial\xi^i}\frac{\partial\chi^\nu_\pm}{\partial\xi^j}.\label{eqj3}
\end{equation}
In (\ref{eqj3}), $n^\pm_\alpha$ denote the components of the normal vector to $\Sigma$ in $\chi^\alpha_{\pm}$ coordinates. 

We express the intrinsic metric on $\Sigma$ as
\begin{equation}
ds^2_{\Sigma} =  d\tau^2-R^2(\tau)(d\theta^2+\sin^2\theta d\phi^2). \label{eqj4} 
\end{equation}
Using the condition (\ref{eqj1}) for the intrinsic metric (\ref{eqj4}) of the interior space-time (\ref{intm1}), we obtain
\begin{eqnarray}
\frac{dt}{d\tau} &=& A(r_{\Sigma},t),\label{bcon1}\\
r_{\Sigma} &=& R(\tau)A(r_\Sigma,t).\label{bcon2}
\end{eqnarray}
In the interior space-time
\begin{equation} 
f(r,t) = r-r_\Sigma = 0.\label{eqj5} 
\end{equation}
The vector with components $\frac{\partial f}{\partial \chi^\alpha_-}$ is orthogonal to $\Sigma$. Accordingly the unit normal to $\Sigma$ in coordinates $\chi^\alpha_-$ is obtained as
\begin{equation} 
n^-_\alpha = \{0,\frac{1}{A\sqrt{1-k r^2}},0,0\}.\label{eqj6}
\end{equation}
The intrinsic curvatures corresponding to the interior region have the
explicit expressions 
\begin{eqnarray}
K^-_{\tau\tau} &=&\left[A'(1-k r^2)^{1/2}\right]_{\Sigma}\label{eqj7}\\
K^-_{\theta\theta} &=&\left[\left(\frac{r}{A}-\frac{r^2 A'}{A^2}\right)(1-k r^2)^{1/2}\right]_{\Sigma}.\label{eqj8}
\end{eqnarray}

Using the condition (\ref{eqj1}) for the metric (\ref{eqj4}) of the exterior space-time (\ref{Vm}), we obtain
\begin{eqnarray}
{\sf r}_\Sigma(v) &=& R(\tau),\label{eqj9}\\
\left(\frac{dv}{d\tau}\right)^{-2}_\Sigma &=& \left(2\frac{d{\sf r}}{dv}+1-\frac{2 m(v)}{\sf r}\right)_\Sigma.\label{eqj10}
\end{eqnarray}
For the exterior region, we have 
\begin{equation}
f({\sf r},v) = {\sf r}- {\sf r}(v) = 0,\label{eqj11} 
\end{equation}
and
\begin{equation}
\frac{\partial f}{\partial\chi^\alpha_+} = \left(-\frac{d{\sf r}}{dv},1,0,0\right).\label{eqj12}
\end{equation} 
The unit normal to $\Sigma$ is  
\begin{equation}
n^+_\alpha = \left(-\frac{d{\sf r}}{d\tau},\frac{dv}{d\tau},0,0\right).\label{eqj13}
\end{equation}
The expressions for the extrinsic curvatures corresponding to the exterior region are found to be
\begin{eqnarray} 
K^+_{\tau\tau} &=& \left[\left(\frac{d^2v}{d\tau^2}\right)\left(\frac{dv}{d\tau}\right)^{-1}-\frac{m}{{\sf r}^2}\frac{dv}{d\tau}\right]_{\Sigma},\label{eqj14}\\
K^+_{\theta\theta} &=& \left[{\sf r}\frac{d{\sf r}}{d\tau} + {\sf r}\frac{dv}{d\tau}\left(1-\frac{2m}{{\sf r}}\right)\right]_{\Sigma}, \label{eqj15}
\end{eqnarray}
where Eq.~(\ref{eqj10}) and its derivative were used to derive Eq.~(\ref{eqj14}). Using Eqs.~(\ref{bcon1}), (\ref{bcon2}), (\ref{eqj9}) and (\ref{eqj10}) in Eqs.~(\ref{eqj8}) and (\ref{eqj15}) and imposing the condition
\begin{equation}
(K^+_{\theta\theta})_\Sigma  = (K^-_{\theta\theta})_\Sigma,\label{eqj16}
\end{equation} 
we obtain
\begin{equation} 
m(v) = \left[\frac{r}{2A}\left[kr^2+2r(1-kr^2)\frac{A'}{A}+r^2\left(\frac{\dot{A}}{A}\right)^2-r^2(1-kr^2)\left(\frac{A'}{A}\right)^2\right]\right]_{\Sigma}.\label{mm}
\end{equation}
It follows from Eq.~(\ref{mm}), that the mass within the sphere of radius $r_\Sigma$ at any instant $t$ can be expressed as
\begin{equation} 
m(r,t) \stackrel{\Sigma}{=}   \frac{r}{2A}\left[kr^2+2r(1-kr^2)\frac{A'}{A}+r^2\left(\frac{\dot{A}}{A}\right)^2-r^2(1-kr^2)\left(\frac{A'}{A}\right)^2\right].\label{mas}
\end{equation}
Using Eqs.~(\ref{Eq3}), (\ref{Eq4}), (\ref{bcon1}), (\ref{bcon2}), (\ref{eqj7}), (\ref{eqj8}), (\ref{eqj9}),  (\ref{eqj14}), (\ref{eqj15}) and  (\ref{mm})
and imposing the condition 
\begin{equation}
(K^+_{\tau\tau})_\Sigma = (K^-_{\tau\tau})_\Sigma,\label{eqj17} 
\end{equation}
we finally obtain
\begin{equation}
p \stackrel{\Sigma}{=} \left[\frac{q}{A\sqrt{1-kr^2}}\right]_\Sigma.\label{eqj18}
\end{equation}

\section{\label{sec4} Behaviour of the physical quantities}
We shall consider the cases $k=0$ and $k \neq 0$ separately.
\subsection{Case I: $k = 0$}
The models of collapsing fluid configurations of this class are characterized by the following expressions for energy-density, pressure and heat flux, respectively:
\begin{eqnarray}
\rho &=& 3\left[4\xi\eta+(r^2\dot{\xi}+\dot{\eta})^2\right], \label{eneq1}\\
p &=& 12r^2\xi^2-8\xi(r^2\xi+\eta)-3(r^2\dot{\xi}+\dot{\eta})^2+2(r^2\xi+\eta)(r^2\ddot{\xi}+\ddot{\eta}),  \label{preseq1}\\
q &=& -4r(r^2 \xi+\eta)^2\dot{\xi}.\label{heq1}
\end{eqnarray}
The inhomogeneities in matter density distribution are observed to be linked with the dissipative processes and disappear when the heat flux vanishes.

The total mass of fluid contained within the spherical region of radius $r_{\Sigma}$ at any instant $t$ has the explicit expression
\begin{equation}
m(r,t)\stackrel{\Sigma}{=}  \frac{r^3[4\xi\eta+(r^2\dot{\xi}+\dot{\eta})^2]}{2(r^2\xi+\eta)^3}.\label{massfn}
\end{equation}
The parameter of expansion is
\begin{equation}
\Theta = u^\alpha_{;\alpha} = - 3\dot{A} = -3(r^2\dot{\xi}+\dot{\eta}).\label{expan1}
\end{equation}
In view of Eqs.~(\ref{preseq1}) and (\ref{heq1}) the condition (\ref{eqj18}), is equivalent to 
\begin{eqnarray}
2(r_{\Sigma}^2\xi+\eta)(r_{\Sigma}^2\ddot{\xi}+\ddot{\eta})-3(r_{\Sigma}^2\dot{\xi}+\dot{\eta})^2+4r_{\Sigma}(r_{\Sigma}^2\xi+\eta)\dot{\xi}\nonumber\\-8\xi(r_{\Sigma}^2\xi+\eta)+12r_{\Sigma}^2\xi^2=0.\label{cons1}
\end{eqnarray} 
This relation governs evolution of the physical quantities. Eq.~(\ref{cons1}) is a highly non-linear equation relating two unknown arbitrary functions ($\xi(t)$, $\eta(t)$). A suitable choice of one of the two functions is essential to determine the other. The functions are further constrained by the requirements of (weak) energy conditions  ($\rho,~p,~q > 0$) and other physical requirements. For a collapsing sphere $\Theta$ should be negative implying $(r^2\dot{\xi}+\dot{\eta}) > 0$. Positive heat flux is ensured if $\dot{\xi} < 0$. From (\ref{eneq1}), we conclude that $d\rho/dr = 12r\dot{\xi}(r^2\dot{\xi}+\dot{\eta})$. Since,  $(r^2\dot{\xi}+\dot{\eta}) > 0$ and $\dot{\xi} < 0$, it follows that $d\rho/dr < 0$. The implications of other physical requirements can be examined only by adopting numerical procedures.

\subsection{Case II: $k \neq 0$}
The energy-density, pressure and heat flux of the collapsing fluid configurations of this class have the following expressions:
\begin{eqnarray}
\rho &=& 3\left[-k\xi^2+k\eta^2+(1-kr^2){\dot{\xi}}^2+ 2\sqrt{1-kr^2}\dot{\xi}\dot{\eta}+{\dot{\eta}}^2\right], \label{eneq2}\\
p &=& [3k^2 r^2 \xi^2-k(\sqrt{1-k r^2}\xi+\eta)^2+4k\xi((1-kr^2)\xi +\sqrt{1-kr^2}\eta)-\nonumber\\
&& 3(\sqrt{1-kr^2}\dot{\xi}+\dot{\eta})^2+2(\sqrt{1-kr^2}\xi+\eta)(\sqrt{1-kr^2}\ddot{\xi}+\ddot{\eta})],\label{preseq2}\\
q &=& 2kr\sqrt{1-kr^2}(\sqrt{1-kr^2}\xi+\eta)^2\dot{\xi}.\label{heq2}
\end{eqnarray}
The presence of dissipation in this case also is observed to give rise to inhomogeneities in matter density distribution of the collapsing fluid.

The total mass of the fluid enclosed within the spherical region of radius $r_{\Sigma}$ of the configuration of this class at any instant $t$ is
\begin{equation}
m(r,t) \stackrel{\Sigma}{=}  \frac{r^3}{2(\sqrt{1-kr^2}\xi+\eta)^3}\left[k(-\xi^2+\eta^2)+(1-kr^2){\dot{\xi}}^2 + 2\sqrt{1-kr^2}\dot{\xi}\dot{\eta}+{\dot{\eta}}^2\right].\label{massfn2}
\end{equation}
The expansion parameter is
\begin{equation}
\Theta = -3(\sqrt{1-kr^2}\dot{\xi}+\dot{\eta}).\label{expan2}
\end{equation}
In view of Eqs.~(\ref{preseq2}) and (\ref{heq2}), the condition (\ref{eqj18}), is equivalent to
\begin{eqnarray}
2(\sqrt{1-kr_{\Sigma}^2}\xi+\eta)(\sqrt{1-kr_{\Sigma}^2}\ddot{\xi}+\ddot{\eta})-2kr_{\Sigma}(\sqrt{1-kr_{\Sigma}^2}\xi+\eta)\dot{\xi} \nonumber\\
+3 r_{\Sigma}^2 k^2 \xi^2 -3(\sqrt{1-kr_{\Sigma}^2}\dot{\xi}+\dot{\eta})^2
-k(\sqrt{1-kr_{\Sigma}^2}\xi+\eta)^2\nonumber\\
+4\xi k[(1-kr_{\Sigma}^2)\xi +\sqrt{1-kr_{\Sigma}^2}\eta]= 0, \label{cons2}
\end{eqnarray}
which is the relation governing evolution of the various physical quantities. Again, one needs to make a suitable choice of one of the two arbitrary functions to determine the other. The functions are further required to comply with the requirements of energy conditions  ($\rho,~p,~q > 0$) and physical plausibility requirements. In this case, $\Theta$ will be negative if $\sqrt{1-kr^2}\dot{\xi}+\dot{\eta} > 0$. Positivity of heat flux is ensured if $\dot{\xi} > 0$ for $k > 0$ and $\dot{\xi} < 0$ for $k < 0$. The implications of other physical requirements can be examined using numerical procedures.

\section{\label{sec5} Particular models}
In view of the complexity of the relation governing the evolution with time of the collapse, we have used two approaches for examining the physical plausibility of the evolution.
\subsection{Approach: I}

The evolution of the collapsing systems is governed by Eqs.~(\ref{cons1}) and (\ref{cons2}). In order to analyze physical viability of the model, it is essential to solve these equations. Note that the metric (\ref{intm1}) is conformal to the Roberson-Walker metric for which we have $k = 0, \pm 1$.  For $k = +1$, we have $0 < r_\Sigma \leq 1$. However, there is no such restriction on $r_\Sigma$ ($0,\infty$) for $k = 0$ and $k=-1$. 

We consider a particular case $k = + 1$ for which the maximum value of the coordinate parameter is $r_\Sigma =1$. Substituting these values in Eq.~(\ref{cons2}), we obtain
\begin{equation} 
2\eta\ddot{\eta} - 2\eta \dot{\xi} -3{\dot{\eta}}^2+3\xi^2-\eta^2 = 0.\label{eqc1}
\end{equation}
We need to specify one of the two unknowns to determine the other. Accordingly, we assume
\begin{equation}
\eta = -\frac{n^2}{t},\label{eqc2}
\end{equation}
where, $n$ is a constant. Solution of Eq.~(\ref{eqc1}) is then obtained as
\begin{equation}
\xi = \frac{n^2\left[e^{\sqrt{3}t}(\sqrt{3}+3t)-3Q(\sqrt{3}t-1)\right]}{3t^2(\sqrt{3}e^{\sqrt{3}t}+3Q)},\label{eqc3}
\end{equation}
where, $Q$ is an integration constant. 

\begin{figure*}
\includegraphics[width=0.75\textwidth]{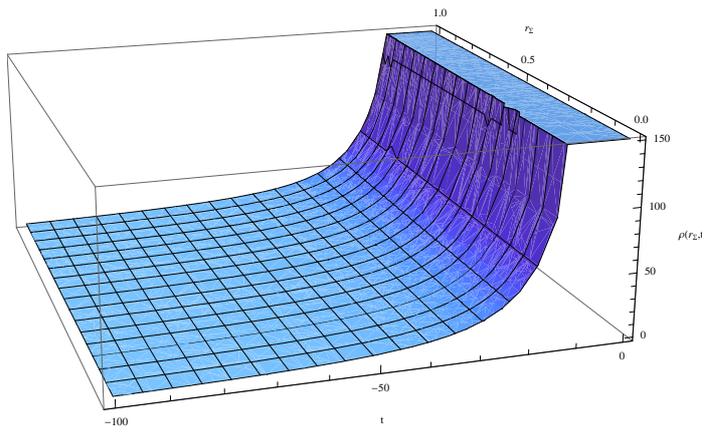}\caption{Evolution of surface density. (We have assumed $k=1$, $n=10$ and $Q=1$.)} \label{fg1}
\end{figure*}

\begin{figure*}
\includegraphics[width=0.75\textwidth]{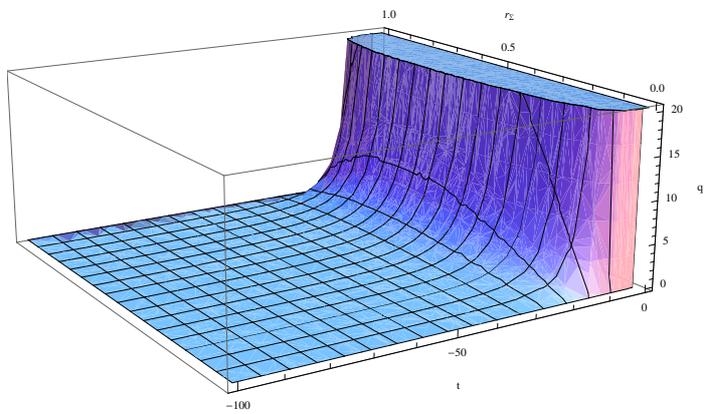}\caption{Evolution of heat-flux. (We have assumed $k=1$, $n=10$ and $Q=1$.)} \label{fg2}
\end{figure*}

\begin{figure*}
\includegraphics[width=0.75\textwidth]{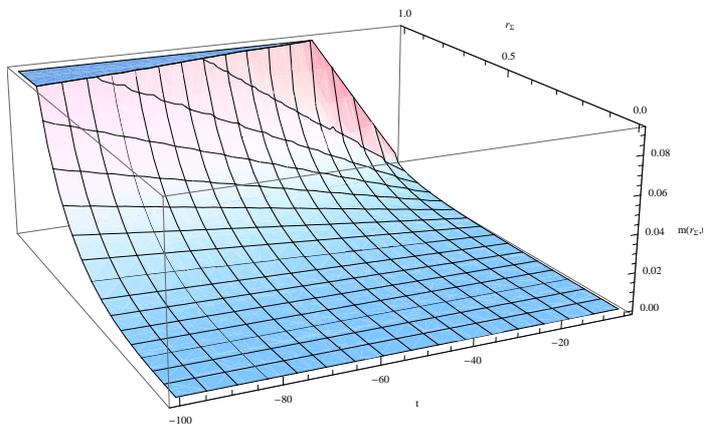}\caption{Evolution of mass. (We have assumed $k=1$, $n=10$ and $Q=1$.)} \label{fg3}
\end{figure*}

\begin{figure*}
\includegraphics[width=0.75\textwidth]{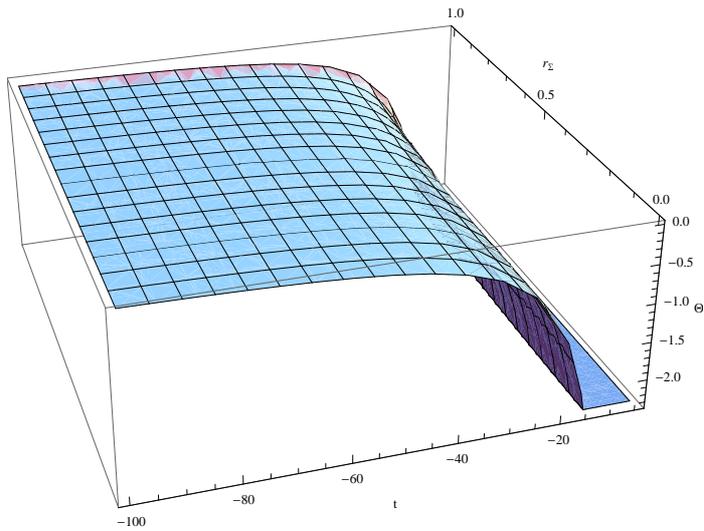}\caption{Rate of expansion. (We have assumed $k=1$, $n=10$ and $Q=1$.)} \label{fg4}
\end{figure*}

Thus, a particular model has been developed which can be utilized to analyze evolution of the physical  parameters. For this particular model, Fig.~\ref{fg1} and \ref{fg2} indicate evolutions of the surface density and heat flux, respectively, beginning with zero values at $t = -\infty$ when $r_\Sigma =1$. Fig.~\ref{fg3} indicates that the mass decreases monotonically as $t \rightarrow 0$. In Fig.~\ref{fg4} the expansion rate ($\Theta$) has been plotted which remains negative throughout the collapse indicating contraction of the collapsing configurations.

\subsection{Approach: II}
In this approach, we adopt an appropriate approximation procedure to examine the collapsing configurations with $k \neq 0$.
We set $r_\Sigma = 1$ in equation Eq.~(\ref{cons2}) rewriting it as
\begin{eqnarray}
2(\sqrt{(1-k)}\xi+\eta)(\sqrt{(1-k)}\ddot{\xi}+\ddot{\eta})-3(\sqrt{(1-k)}\dot{\xi}+\dot{\eta})^2\nonumber\\
-2k(\sqrt{(1-k)}\xi+\eta)\dot{\xi}+4\xi k((1-k)\xi+\sqrt{(1-k)}\eta)\nonumber\\
-k(\sqrt{(1-k)}\xi+\eta)^2+3 k^2 \xi^2 = 0.\label{cons3}
\end{eqnarray}
Since a closed form solution of Eq.~(\ref{cons3}) is not available, we generate an approximate solution by assuming
\begin{equation}
\eta(t)=P e^{n t}, ~~~\xi(t) = \epsilon h(t),\label{eta}
\end{equation}
with $0 < \epsilon << 1$. By considering terms up to $\cal{O}$$(\epsilon)$, an approximate solution of the Eq.~(\ref{cons3}) is obtained in the form
\begin{eqnarray}
\xi(t) &=& -\frac{e^{n t} \sqrt{(1-k)}(k+n^2)P}{2[k^2+n^2+k(-1+n(\sqrt{(1-k)}-n)]}\nonumber\\
&&+  \epsilon\left[C e^\frac{(k+3 n\sqrt{(1-k)}-L)t}{2\sqrt{(1-k)}}+ D e^\frac{(k+3 n\sqrt{(1-k)}+L)t}{2\sqrt{(1-k)}}\right],\label{xi}
\end{eqnarray}
where
$$L = \sqrt{k (-4 + 5 k) + 6 \sqrt{(1 - k)} k n - 5 (-1 + k) n^2},$$
and $P$, $n$, $k$, $C$, $D$ are constants which should be fixed based on physical requirements. 

\begin{figure*} 
\includegraphics[width=0.75\textwidth]{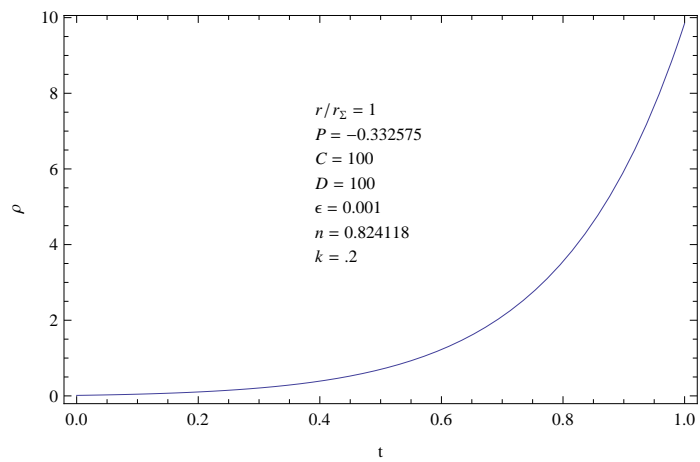}\caption{Evolution of surface density.} \label{fg5}
\end{figure*}

\begin{figure*}
\includegraphics[width=0.75\textwidth]{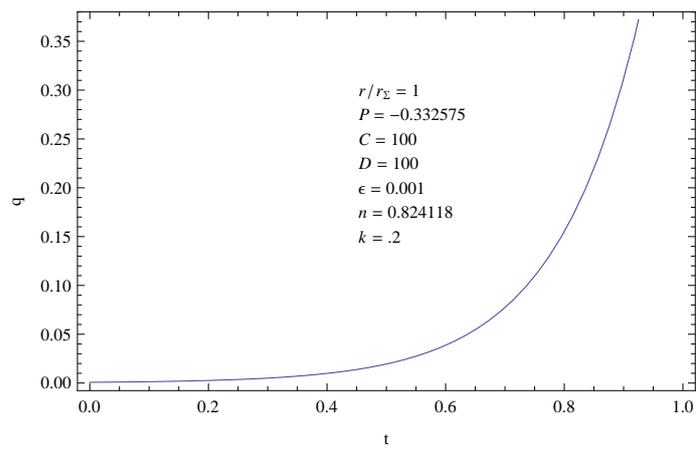}\caption{Evolution of heat-flux.} \label{fg6}
\end{figure*}

\begin{figure*}
\includegraphics[width=0.75\textwidth]{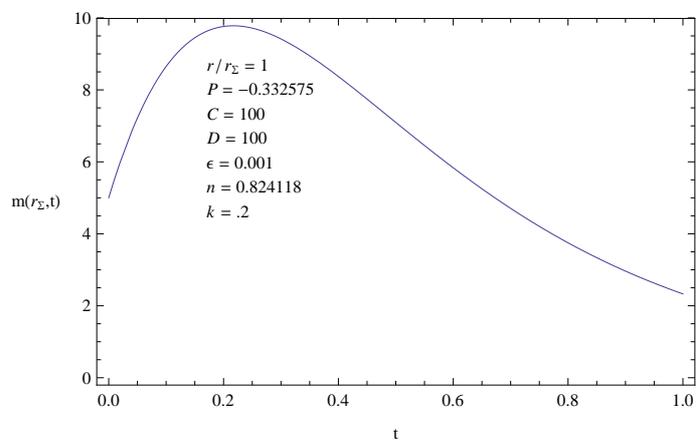}\caption{Evolution of mass.} \label{fg7}
\end{figure*}

\begin{figure*}
\includegraphics[width=0.75\textwidth]{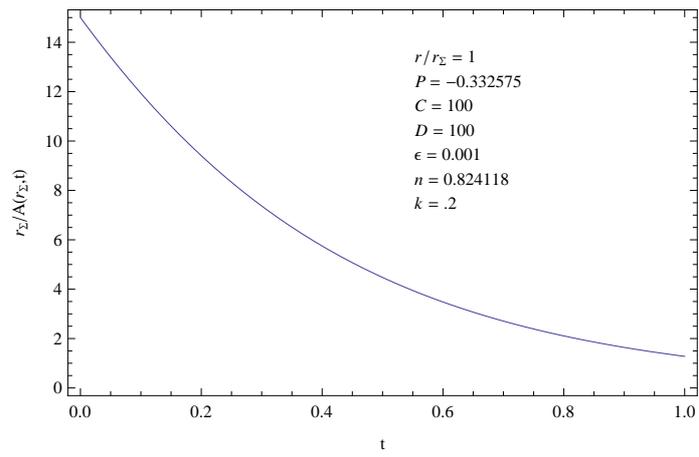}\caption{Evolution of proper radius.} \label{fg8}
\end{figure*}

The behaviour of the physical quantities of the collapsing configuration for this particular model has been examined for configurations with the initial mass and radius $5~M_{\odot}$ and $15~$km, respectively.  The evolution of the physical quantities for specific choices of parameters has been described graphically.
Fig.~\ref{fg5} shows that the surface density $\rho(r_\Sigma, t)$ starts from a finite value and increases as time evolves. The heat-flux $q$ is zero initially and remains positive as time evolves (Fig.~\ref{fg6}). The total mass $m(r_\Sigma, t)$ and geometric radius $(r/A(r,t))_\Sigma$ decrease from their initial values as time progresses as shown in Fig.~\ref{fg7} and \ref{fg8},  respectively.

\section{\label{sec6} Discussions}
  
In this work, we have described relativistic conformally flat solutions for spherically symmetric fluid configurations, undergoing dissipative collapse. It is not difficult to find a similarity between these solutions and the solutions obtained earlier by Herrera {\em et al}\cite{Herrera1} which follow by adopting a different procedure. In the Herrera {\em et al}\cite{Herrera1} approach, conformal flatness is ensured by equating the Weyl tensor of the space-time to zero. The choice of the space-time in our approach is such that this condition is fulfilled on adopting a space-time with geometry conformal to a conformally flat space-time. The RW spacetime is conformally flat and further characterized by the feature that its spatial sections are homogeneous. Accordingly the space-time coordinates of the collapsing configuration are such that the spatial geometry is conformally homogeneous and continues to be so throughout collapse. This approach brings out the nature of impact of the spatial geometry determined by the curvature parameter $k$ on the evolution of the collapse in the presence of dissipation. Further the approach also indicates the inter-dependence between the inhomogeneities in the matter density distribution and presence of dissipative processes during the evolution of the collapse. The mass of the particular collapsing configuration of case $k = 0$ of Approach I, tends to zero as collapse progresses indicating that all energy is radiated away as heat as in a case indicated earlier by Banerjee {\em et al}\cite{Banerjee2}. However, in the case discussed in Approach II, it is not so.

To summarize, the class of conformally flat solutions reported in this article have been utilized to generate physically viable models of spherically symmetric gravitationally collapsing systems dissipating energy in the form of radial heat flux. As for the future investigations, since Maxwell's equations are conformally invariant, it would be interesting to incorporate charge into our model and investigate the impact of electromagnetic field on the overall evolution of the system.  This issue will be taken up elsewhere.

\begin{acknowledgements}
RS gratefully acknowledges support from the Inter-University Centre for Astronomy and Astrophysics (IUCAA), Pune, India, where a part of this work was carried out under its Visiting Research Associateship Programme. The authors would like to thank the anonymous referee for useful comments on the manuscript.
\end{acknowledgements}

\newpage



\end{document}